# A New Journey to the Center of the Earth

It was not only Jules Verne, author of this famous science fiction novel from 1864, who was fascinated by the thought of what could be in the profound depth below us. Certainly similar questions were asked by people millenia ago. Where does the energy of volcanic eruptions and devastating earthquakes come from? How did our Earth come to existence? And today, it is still the same curiosity that drives us to learn more about our blue planet, whose history dates back 4.5 billion years ago. In this article we explain how neutrino physics can contribute in this effort: we present the latest geoneutrino measurements [1] from the Borexino experiment located deep underground below the Gran Sasso mountains in Italy.

**1) The Earth and its energy budget**
The Earth irradiates 47 ± 2 terawatts [2] of thermal radiation, about 40 times more heat than the energy produced by 440 world nuclear power plants. This was estimated based on temperature gradients measured along thousands of boreholes. Part of this energy is the residual heat from the formation of our planet in the process of accretion of smaller irregular objects. The Earth then underwent differentiation into a layered structure. The metallic core was the first to separate from the primitive mantle, the so called *Bulk Silicate Earth (BSE)*. The latter further differentiated into the present-day, nearly 3000 km thick mantle and the crust. The heterogeneous *Continental Crust (CC)* has a long history and is up to 70 km thick, while the younger *Oceanic Crust (OC)* is much thinner and simpler. The constant cooling of the Earth is balanced by the production of radiogenic heat released in the radioactive decays of long-lived isotopes inside the Earth. The radiogenic heat of the lithosphere, i.e. the crust and the uppermost brittle *Continental Lithospheric Mantle (CLM)*, is relatively well known ($8.1^{+1.9}_{-1.4}$ TW), while the radiogenic heat of the deeper mantle is poorly constrained.

**2) Geoneutrinos: messengers from the deep Earth**
Geoneutrinos are electron flavour antineutrinos emitted inside the Earth, in the radioactive decays of *Heat Producing Elements (HPEs)* with lifetimes compatible with the age of the Earth, such as $^{232}$Th, $^{235}$U, $^{238}$U, and $^{40}$K. The ratio between the number of emitted geoneutrinos and the amount of radiogenic heat is fixed and well known. Naturally, the amount of the Earth's radiogenic heat is then directly proportional to the abundances of *HEPs* inside the Earth. Geoneutrinos are the only direct probe to track the Earth's radiogenic heat: about a million of geoneutrinos are crossing your fingertip each second. Their probability to interact with matter is very small and thus their detection is extremely challenging.

**3) Borexino: a unique instrument to measure geoneutrinos**
Borexino detector is the world's most radio-pure liquid scintillator detector placed at the Laboratori Nazionali del Gran Sasso in Italy at a depth of 3800 m w.e. (meter water equivalent). Borexino overcame its original goal to measure solar neutrinos from $^7$Be decays via neutrino-electron elastic scattering. A comprehensive measurement of *pp*-chain solar neutrinos has been presented recently [3], while the ability to detect antineutrinos was also proven. Borexino observed geoneutrinos for the first time in 2010 [4] and then improved the significance of this measurement to more than 5$\sigma$ in 2015 [5]. Radiopurity of the detector, its calibration, stable performance, relatively large distance to nuclear reactors, as well as the depth of LNGS to guarantee smallness of cosmogenic background, are the main building blocks for a geoneutrino measurement with systematic uncertainty below 5%. This was presented in a new comprehensive analysis [1].

Borexino detector is schematically shown in Fig. 1. Nominally 278 ton of liquid scintillator (pseudocumene (PC, $C_9H_{12}$) with the addition of PPO at 1.5 g/l), held inside the inner vessel (IV), a very thin (125 $\mu$m) nylon balloon, serve as a target for neutrino interactions. The scintillator is shielded from external background by 2.6 m thick layer of PC-based non-scintillating buffer liquid enclosed in a stainless steel sphere of 6.85 m radius. On the latter are mounted 2212 8'' photomultipliers (PMTs) detecting the scintillation light. The buffer liquid is divided in two sub-volumes by an additional nylon vessel blocking the inward diffusion of Radon. The whole structure is placed inside a dome-like steel water tank of 9 m base radius and 16.9 m height filled with approximately 1 kton of ultrapure water. This volume provides additional shielding against gammas and neutrons and, equipped with 208 PMTs, serves also as an active Cherenkov veto of the residual cosmic muons. The triggering threshold is about 50 keV. With the light yield of about 500 photoelectrons (p.e.) per 1 MeV, the energy and position reconstructions' resolutions are about 5% and 10 cm at 1 MeV, respectively. Borexino scintillator is also an excellent medium for $\alpha/\beta$ pulse shape discrimination. The Geant-4 based Monte Carlo (MC) simulation [7] was tuned on an extensive calibration campaign [6] with radioactive sources.

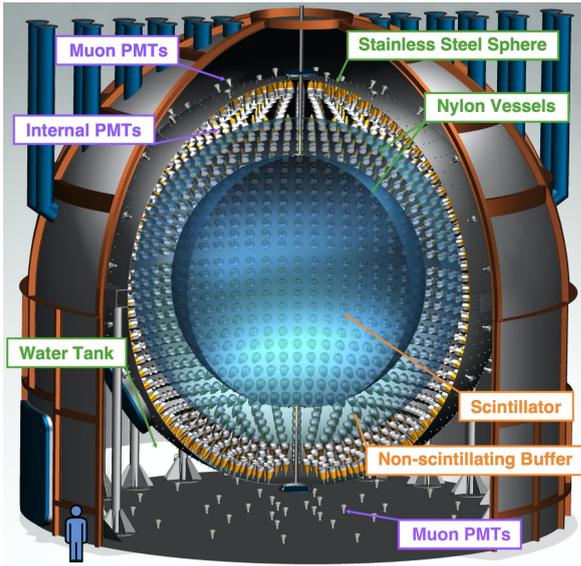

Figure 1: Scheme of the Borexino detector.

### 4) Geoneutrino detection

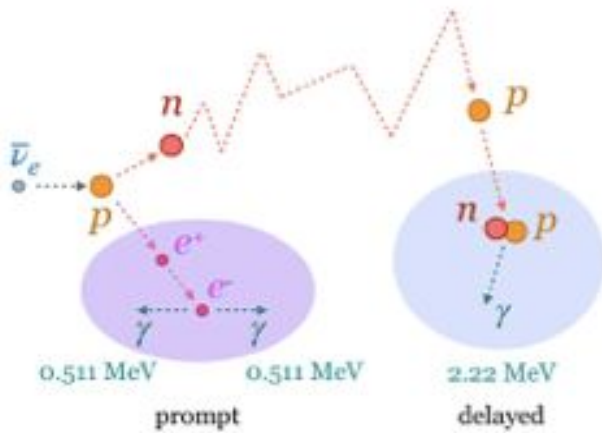

Figure 2: Schematic of the proton Inverse Beta Decay interaction, used to detect geoneutrinos, showing the origin of the prompt (violet area) and the delayed (blue area) signals.

Geoneutrinos are detected via the *Inverse Beta Decay (IBD)* interaction shown in Fig. 2, in which they interact with free protons of the scintillator. The IBD, $\bar{\nu}_e + p \rightarrow e^+ + n$, yields a positron and a neutron. Since the produced neutron is heavier than the target proton, the interaction has a kinematic threshold of 1.8 MeV, which means that we can only detect geoneutrinos coming from the $^{238}$U and $^{232}$Th decay chains. For the sake of comparison either between experiments or between the expected and measured antineutrino signals, the *Terrestrial Neutrino Units (TNU)* is introduced. One TNU corresponds to 1 antineutrino event detected via IBD over 1 year by a detector with 100% detection efficiency containing $10^{32}$ free target protons (roughly 1 kton of liquid scintillator).

The positron promptly comes to rest and annihilates producing two 511 keV gammas. The energy of this *prompt* signal is directly correlated with the incident geoneutrino energy. The neutron, after some time, gets thermalized and is captured on a proton (or with about 1% probability on $^{12}$C) to produce a 2.22 (4.95) MeV ɣ. These gammas interact via several Compton scatterings and give rise to the *delayed* signal. The space and time coincidence between the *prompt* and *delayed* signals provides a golden channel to detect geoneutrinos and to strongly suppress different kinds of backgrounds.

A set of the selection cuts was optimized to maximize the sensitivity to geoneutrinos. These include energy intervals of *prompt* and *delayed*, time interval and spatial distance between them, fiducial volume, and pulse-shape discrimination on *delayed* to eliminate Rn-correlated background. All IBD candidates have to pass the vetoes applied to tag the muons and the muon-induced cosmogenic background. Once a muon passes the detector, four different kinds of complicated space-time vetoes are applied depending on the type of muon.

The most important background for geoneutrino measurement is represented by antineutrinos from the nuclear power plants. It can be evaluated with few % precision using the information provided by the International Atomic Energy Agency. The important non-antineutrino backgrounds that mimic IBD coincidences, namely (i) residual cosmogenic $^9$Li, decaying in electron + neutron, (ii) accidental background between uncorrelated events, and (iii) (α, n) background caused by the alphas from $^{210}$Po in the scintillator, can be evaluated with sufficient precision based on independent data and/or MC simulation.

## 5) Expected geoneutrino signal at LNGS

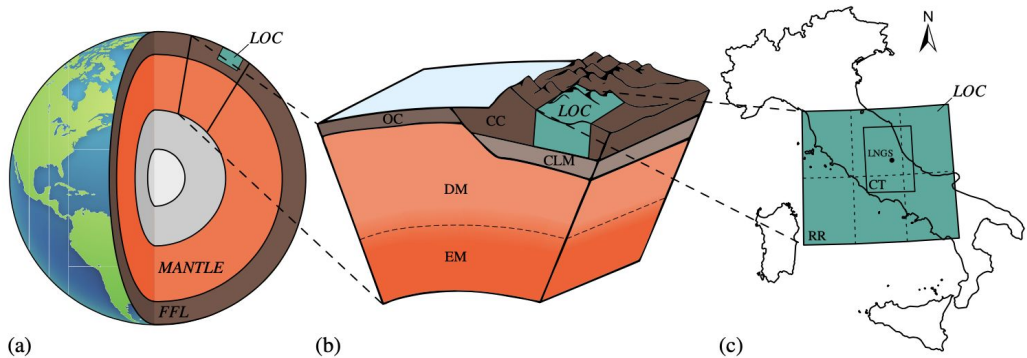

*Figure 3: (a) Schematic drawing of the Earth's with the three components of the expected geoneutrino signal at LNGS: (i) the local crust (LOC),* 492 x 444 km area around the detector, *(ii) the Far Field lithosphere (FFL), and (iii) the mantle. The metallic core (in grey) is not expected to contain U and Th. Not in scale. (b) Detailed view (not in scale) of a section of the silicate Earth: continental crust (CC), oceanic crust (OC), continental lithospheric mantle (CLM), enriched mantle (EM), and depleted mantle (DM). (c) Simplified map of the LOC.*

The flux of geoneutrinos reaching the detector is estimated using the information about *HPEs'* isotopic abundances in the rocks inside the Earth and the known antineutrino production rate per unit mass of *HPEs,* integratedthe over the entire volume of the Earth. In order to convert the flux to an expected signal in TNU units, one has to take into account the electron-flavour survival probability (IBD interaction is sensitive only to $\overline{\nu}_e$ ), the energy spectrum of emitted antineutrinos, the IBD interaction cross section as a function of energy, and the number of target protons.

In the process of integration, three different levels of precision concerning the *HPEs* distributions are adopted. The local crust (*LOC*) (Fig. 3) contribution (~40%) is estimated based on the refined geological model [9]. The contribution from the *Far-Field Lithosphere* (*FFL,* Fig. 3) can be calculated relatively precisely adopting the 1° × 1° geophysically based, 3D global reference model [10]. The estimated geoneutrino signal from the bulk lithosphere (*LOC* + *FFL)* is $25.9^{+4.9}_{-4.1}$ TNU. The *HPEs'* masses in the mantle are taken from a range of *BSE* models, after the subtraction of the *HPEs'* masses known to be in the lithosphere, and their distribution across the mantle is then varied within some reasonable assumptions. The total expected geoneutrino signal at LNGS can then vary widely between $28.5^{+5.5}_{-4.8}$ TNU and $45.6^{+5.6}_{-4.9}$ TNU, depending on the prediction of mantle signal from different Earth models.

**6) Spectral analysis in a nutshell**

After a selection of IBD candidates with optimized cuts, the geoneutrino signal is extracted from the unbinned-likelihood spectral fit of the charges of all *prompts*. The shapes of all spectral components used in such a fit are taken from the MC-constructed *Probability Distribution Functions* (*PDFs*), with the exception of the accidental background, which can be measured with sufficient precision by off-time coincidence. The result of the fit is the number of events due to each spectral component. Naturally, the number of geoneutrinos is kept free in the fit, as well as that of reactor antineutrinos. The number of events due to non-antineutrino backgrounds is however constrained to the values obtained from the analyses of independent data. With a known exposure and detection efficiency, the number of detected geoneutrinos and reactor antineutrinos can be expressed in the units of TNU.

**7) Borexino latest results**

In the period between December 9, 2007 and April 28, 2019, corresponding to 3262.74 days of data acquisition, 154 IBD candidates [1] were found. The exposure of $(1.29 \pm 0.05) \times 10^{32}$ protons × year represents an increase by a factor of two over a previous analysis [6]. For geoneutrinos, the $(87.0 \pm 1.5)$% detection efficiency was estimated with MC. The time, spatial, and energy distributions of the IBD candidates follow the expectations. In the fit, the geoneutrino component was first modelled with a PDF, in which the ratio of $^{232}$Th and $^{238}$U signals was fixed to 0.27. The latter value corresponds to a chondritic ratio of Th and U masses of 3.9, expected to hold for the bulk Earth. By observing $52.6^{+9.4}_{-8.6}$ (stat) $^{+2.7}_{-2.1}$ (sys) geoneutrinos from $^{238}$U and $^{232}$Th, a geoneutrino signal of $47.0^{+8.4}_{-7.7}$ (stat) $^{+2.4}_{-1.9}$ (sys) TNU was obtained. The $^{+18.3}_{-17.2}$ % total precision is in agreement with the expected sensitivity based on generation of 10,000 pseudo-experiments. The observed geoneutrino signal is compatible with the geological models, with a preference for those predicting high U and Th concentrations. The resulting reactor antineutrino signal $80.5^{+9.8}_{-9.3}$ (stat) $^{+4.1}_{-4.4}$ (sys) TNU is compatible with the expectations of $84.5^{+1.5}_{-1.4}$ TNU and $79.6^{+1.4}_{-1.3}$ TNU (calculated without and with consideration of the 5 MeV excess observed by reactor antineutrino experiments, respectively), which confirms the Borexino precision to detect antineutrinos. The systematic uncertainties include contributions from atmospheric neutrino background, uncertainty in the shape of the reactor antineutrino spectrum, error on MC efficiency, as well as position and IV-shape reconstructions. Compatible results were found when contributions from $^{238}$U and $^{232}$Th were both fit as free parameters. Unfortunately, with an achievable exposure, Borexino does not have any sensitivity to determine U/Th ratio of the geoneutrino signal.

The mantle signal was extracted from the spectral fit (Fig. 4) by constraining the contribution from the bulk lithosphere. In the lithospheric PDF, the $^{232}$Th and $^{238}$U signals were scaled with Th/U signal ratio of 0.29, that is based on geological observations. Instead, in the mantle PDF, the applied Th/U signal ratio was 0.26. This procedure maintains the chondritic Th/U mass ratio for the bulk Earth . The measured mantle signal is $21.2^{+9.5}_{-9.0}$ (stat) $^{+1.1}_{-0.9}$ (sys) TNU and the null-hypothesis of observing mantle geoneutrinos is excluded at a 99.0% C.L. This measurement constrains at 90% C.L. a mantle composition with abundance of Uranium > 13ppb and that of Thorium >48 ppb, assuming for the mantle homogeneous distribution of U and Th and a Th/U mass ratio of 3.7.

The measured mantle signal can be converted to the radiogenic heat from U and Th in the mantle of $24.6^{+11.1}_{-10.4}$ TW. Assuming 18% contribution of $^{40}$K in the total mantle radiogenic heat and $8.1^{+1.9}_{-1.4}$ TW of total radiogenic heat of the lithosphere, the Borexino estimate of the total radiogenic heat of the Earth is $38.2^{+13.6}_{-12.7}$ TW (black point in Fig. 5), which corresponds to the convective Urey ratio of $0.78^{+0.41}_{-0.28}$. Figure 5 demonstrates the Earth heat as predicted by different *BSE* models and in the last vertical bar, as estimated from the Borexino measurement of the mantle geoneutrino signal. The lithospheric radiogenic heat is shown in brown and is the same for all vertical bars. In orange is shown the radiogenic heat from the mantle, the prediction of which largely varies for different *BSE* models. In the last bar, the orange colour represents the Borexino estimation of the mantle radiogenic heat. The blue color represents the heat from the Earth's secular cooling, that fills up the "available gap" between the total radiogenic heat and the (47 ± 2) TW of the measured total Earth surface heat flux (represented in Fig. 5 by the horizontal band), as an upper limit for the Earth's heat decomposed into the above mentioned 3 main categories. The mantle radiogenic heat estimated by Borexino is compatible with different geological predictions, however there is about 2.4σ tension with those Earth models which predict the lowest *HPE*s' concentration in the mantle. Borexino estimates 85% probability that the radioactive decays produce more than half of the total terrestrial heat.

With the application of a constraint on the number of expected reactor antineutrino events, Borexino has placed an upper limit on the number of events from a hypothetical Uranium-fission georeactor [9] inside the Earth. Borexino excludes the existence of a georeactor with a power greater than 0.5/2.4/5.7 TW at 95% C.L., assuming its location at 2900/6371/9842 km distance from the detector.

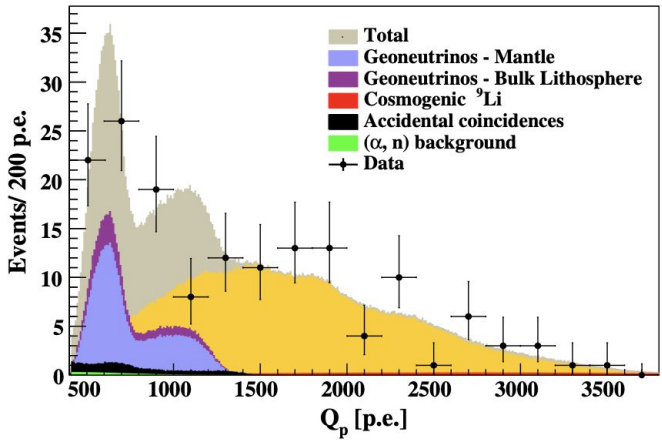

*Figure 4: Borexino spectral fit of 154 IBD candidates to extract the mantle signal after constraining the contribution of the bulk lithosphere [1].*

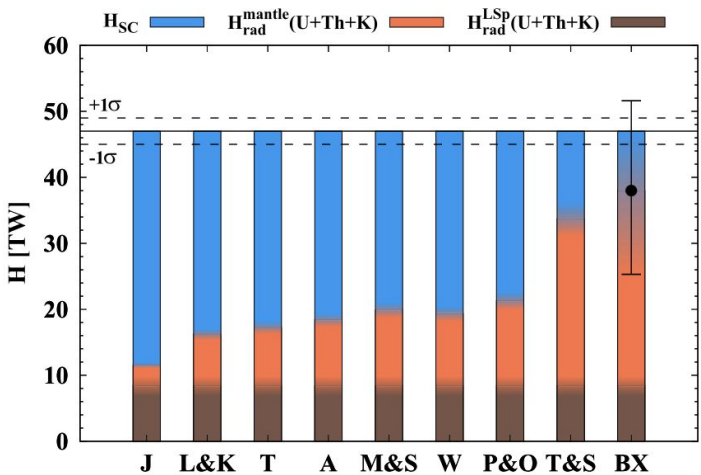

*Figure 5: The Earth heat for different BSE models and in the last column, for the Borexino data point [1]. Details in text.*

## 8) Perspectives of Neutrino Geoscience

Neutrino Geoscience is a newly born interdisciplinary field having as its main aim determination of the Earth's radiogenic heat, a key parameter in understanding of our planet. The existing geoneutrino measurements, as the one of Borexino [1] presented here or that of KamLAND experiment in Japan [11], has proven that with the current technology we are able to measure geoneutrinos and extract the first geological implications. The existing measurements also confirm the general validity of different geological models predicting the U and Th abundances in the Earth. This is an enormous success of both neutrino physics and geosciences. However, additional and more precise measurements are needed in order to extract firm geological results. SNO+ [12] in Canada is currently filling its detector with liquid scintillator and has geoneutrinos among its physics goals. The 20 kton JUNO [13] is under construction and will be operational in 2022. Under consideration is also a detector in the world's deepest laboratory JINPING in China [14]. HanoHano [15] is an interesting proposal to use a movable 5 kton detector resting on the ocean floor. As the oceanic crust is particularly thin and relatively depleted in *HPEs*, this experiment could provide the most direct information about the mantle. It is anticipated that using antineutrinos to study the Earth's interior will increase in the future based on the availability of new detectors and the continuous development of analysis techniques.

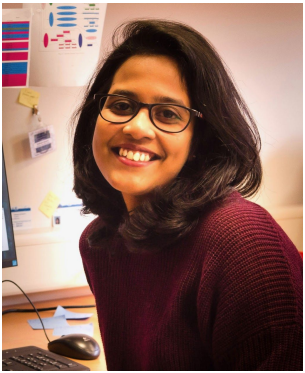

S. Kumaran
Institut für Kernphysik IKP-2, Forschungszentrum Jülich, 52425 Jülich, Germany
III. Physikalisches Institut B, RWTH Aachen University, 52062 Aachen, Germany

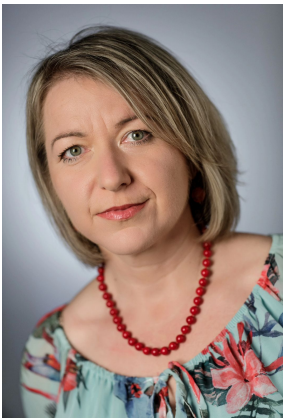

L. Ludhova
Institut für Kernphysik IKP-2, Forschungszentrum Jülich, 52425 Jülich, Germany
III. Physikalisches Institut B, RWTH Aachen University, 52062 Aachen, Germany